\documentclass[10pt]{article}

\usepackage[aux]{rerunfilecheck}
\usepackage{amsmath}
\usepackage{amssymb}
\usepackage{bm}
\usepackage{mathtools}
\usepackage[numbers,sort&compress]{natbib}
\usepackage[autostyle]{csquotes}
\usepackage{graphicx}
\usepackage[usenames,dvipsnames]{color}
\usepackage{bbm} % needed for \mathbbm{1}
\usepackage{placeins} % Needed for \FloatBarrier
\usepackage{booktabs}
\usepackage[labelfont=bf]{caption}
\usepackage[top=1in, bottom=1.25in, left=0.9in, right=0.9in]{geometry}
\usepackage{subfigure}
\usepackage{tabularx}
\usepackage{multirow}
\usepackage[hidelinks]{hyperref}
\usepackage{braket}

\title{\bf Addendum: Fitting the DESI BAO Data with Dark Energy Driven by the Cohen--Kaplan--Nelson Bound}

\author{Patrick Adolf$^1$\footnote{patrick.adolf@tu-dortmund.de},
    Martin Hirsch$^2$\footnote{mahirsch@ific.uv.es},
    Sara Krieg$^1$\footnote{sara.krieg@tu-dortmund.de},
    Heinrich P\"as$^1$\footnote{heinrich.paes@tu-dortmund.de},
    Mustafa Tabet$^1$\footnote{mustafa.tabet@tu-dortmund.de}
\smallskip
\\
{\it $^1$Fakult\"at f\"ur Physik,
Technische Universit\"at Dortmund, D-44221 Dortmund, Germany} \\
{\it $^2$Instituto de F\`isica Corpuscular (IFIC), Universidad de Valencia-CSIC,}\\
{\it E-46980 Valencia, Spain}
}

\begin{document}

\maketitle

% ##################################################################################
% #################################### Abstract ####################################
% ##################################################################################
\begin{abstract}
  Motivated by the recent Year-2 data release of the DESI collaboration, we
  update our results on time-varying dark energy models driven by the
  Cohen--Kaplan--Nelson bound.
  The previously found preference of time-dependent dark energy models compared
  to $\Lambda$CDM is further strengthend by the new data release.
  For our particular models, we find that this preference increases up to
  $\approx 2.6\,\sigma$ depending on the used supernova dataset.
\end{abstract}

% ######################################################################################
% #################################### Introduction ####################################
% ######################################################################################
\section{Introduction}
In this addendum, we update the results of our previous
work~\cite{Adolf:2024twn} in the light of the recent Year-2 data release (DR2)
of the Dark Energy Spectroscopic Instrument (DESI)
collaboration~\cite{DESI:2025zgx}.
We consider a dark energy scaling proportional to the squared Hubble parameter
$H(z)$ as
\begin{equation}\label{eq:ckn}
  \rho_\text{DE}(z) = \Lambda_0 + \nu \frac{M_\text{Pl}^2 H^2(z)}{16\pi^2}\,,
\end{equation}
where $\Lambda_0$ is the contribution of the classical cosmological constant to
the dark energy density $\rho_\text{DE}(z)$, $M_\text{Pl}$ is the Planck mass,
and $\nu$ is a free parameter of the model.
As discussed in our previous work, this scaling is motivated by the
Cohen--Kaplan--Nelson (CKN) bound~\cite{Cohen:1998zx}, used to connect the
quantum corrections of the dark energy density to the size of the universe.
Hence, as introduced in Reference~\cite{Adolf:2024twn}, we call the dark energy
density evolving from Equation~\eqref{eq:ckn} $\nu$CKN model, or simply CKN model for
the case $\nu=1$.

In Section~\ref{sec:results}, we show the result of our combined analysis of the
DESI baryonic acoustic oscillations (BAO) DR2~\cite{DESI:2025zgx}, supernova
datasets~\cite{Brout:2022vxf,DES:2024tys} and model-independent Hubble
measurements~\cite{Favale:2024lgp,Moresco:2020fbm}.
We compare the results of the ($\nu$)CKN models with other dark energy models in
the literature, and discuss the difference to the fit for which the DESI BAO
Year-1 data release (DR1)~\cite{DESI:2024mwx} is used instead.
Finally, we summarize and discuss the results in Section~\ref{sec:discussion}.

% ######################################################################################
% #################################### Results #########################################
% ######################################################################################
\section{Updated Results}
\label{sec:results}
Analogously to the statistical procedure described in
Reference~\cite{Adolf:2024twn}, we perform a $\chi^2$ fit and provide the
resulting best-fit points of the CKN and $\nu$CKN model in
Table~\ref{tab:CKN_with_data}.
Since we use two different supernova datasets, DES-SN5YR (DESY5) and Pantheon+,
we always combine the DESI BAO DR2 and Hubble data with one of them separately.
The resulting $\chi^2$ values show that both the CKN and the more general $\nu$CKN models are well
compatible with the experimental measurements, as the goodness of the best-fit
points over the degrees of freedom (DOF) are $\chi^2/\text{DOF} \approx 0.89$
and $\chi^2/\text{DOF} \approx 0.88$ for the DESY5 and Pantheon+ data,
respectively.

\begin{figure}
  \centering
  \begin{minipage}{0.45\textwidth}
    \includegraphics[scale=0.9]{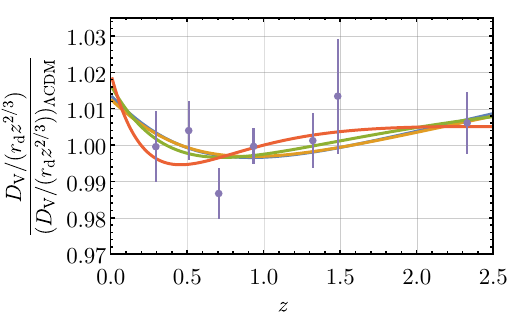}
  \end{minipage}
  \begin{minipage}{0.45\textwidth}
    \includegraphics[scale=0.9]{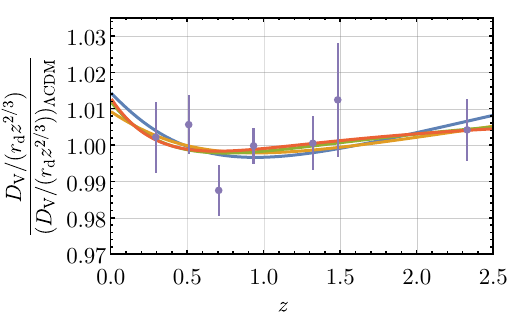}
  \end{minipage}
  \includegraphics[scale=0.9]{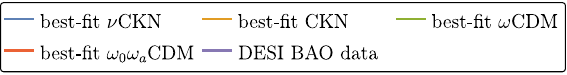}
  \caption{Shown is the angle-averaged distance quantity $D_\text{V}/(r_\text{d}
  z^{2/3})$ normalised to the $\Lambda$CDM value at our best-fit point
  together with the DR2 DESI measurements~\cite{DESI:2025zgx} for all considered
  dark energy models. In the left plot, the best-fit points of the combination
  with the DESY5 dataset
  are used and in the right plot with the Pantheon+ dataset.
  Note that in the left plot the best-fit lines for CKN and $\nu$CKN overlap.}
  \label{fig:DESI}
\end{figure}

For the comparison to other dark energy models, we provide the best-fit
points of $\Lambda$CDM, $\omega$CDM, and
$\omega_0\omega_a$CDM~\cite{Chevallier:2000qy,Linder:2002et} in
Table~\ref{tab:LCDM_with_data} and their $\chi^2$ difference to the ($\nu$)CKN
models in Table~\ref{tab:CKN_with_models}.
To compare between models with a different number of model parameters $k$, we
use the Akaike information criterion (AIC)
\begin{align}
  \text{AIC} = \chi_\text{min}^2 + 2k\,.
\end{align}
The results show that the CKN and $\nu$CKN models are preferred with respect to
the $\Lambda$CDM model for both datasets.
In case of the $\nu$CKN model, the
$\chi^2$ difference can be translated into a significance of $2.63\,\sigma$ and
$1.75\,\sigma$ for the DESY5 and the Pantheon+ data, respectively.
However, the $\Delta \chi^2$ and $\Delta$AIC values show that the $\omega$CDM
and $\omega_0\omega_a$CDM models provide an even better fit to the data.
Only according to the $\Delta$AIC values for the combination with the Pantheon+ dataset, both, the CKN and $\nu$CKN
models are slightly preferred with respect to the $\omega_0\omega_a$CDM model.
The prediction of the angle-averaged distance quantity $D_\text{V}/(r_\text{d}z^{2/3})$ at our
best-fit points normalised to the $\Lambda$CDM prediction is shown in
Figure~\ref{fig:DESI} together with the DESI DR2 measurements.
The differences between the $\chi^2$ values of the DESI BAO DR1 and DR2 are
presented in Table~\ref{tab:Change} and demonstrate that the fit of all models
improved.
However, the improvement of the fit for the $\Lambda$CDM model is
non-significant, whereas the new measurements have a higher impact on the
time-dependent dark energy models.
In the case of the Pantheon+ data, the improvement of the $\Lambda$CDM model is
larger than for the DESY5 data. The most significant change can be found for the
$\omega$CDM and $\nu$CKN models, for the DESY5 and Pantheon+ supernova dataset,
respectively.

\begin{figure}
  \includegraphics[scale=1]{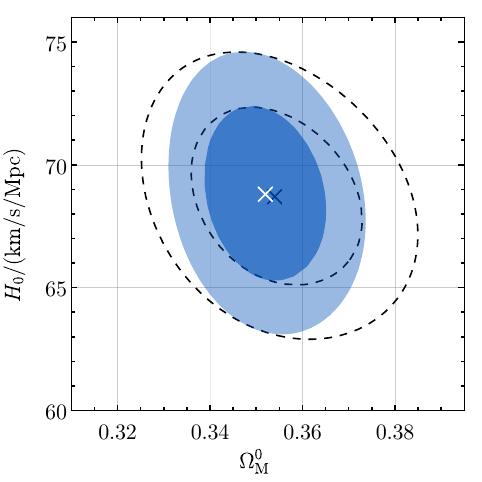} 
  \includegraphics[scale=1]{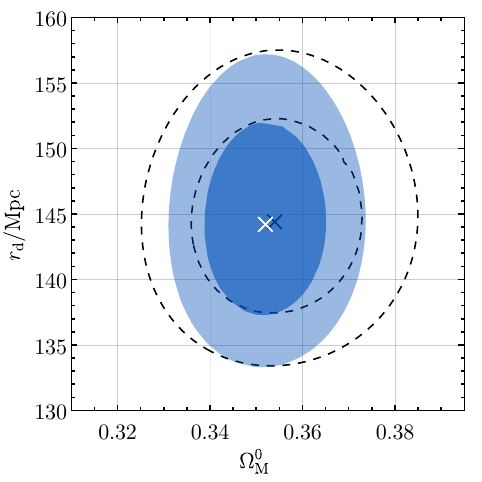}
  \includegraphics[scale=1]{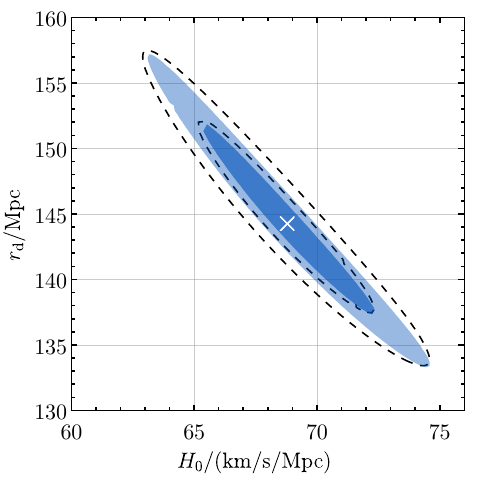} \hspace*{10em}
  \raisebox{10em}{\includegraphics[scale=1]{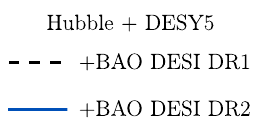}}
  \caption{Correlations of $H_0$--$\Omega_\text{M}^0$ (top left),
    $H_0$--$r_\text{d}$ (top right), $r_\text{d}$--$\Omega_\text{M}^0$ (bottom
    left) in the CKN model for the DESI BAO+Hubble+DESY5 DR1 (black dashed lines)
    and DESI BAO+Hubble+DESY5 DR2 (blue area) dataset at the $95\,\%$ and $68\,\%$ CL.}
  \label{fig:Corr_CKN_DESY5}
\end{figure}

\begin{figure}
  \includegraphics[scale=1]{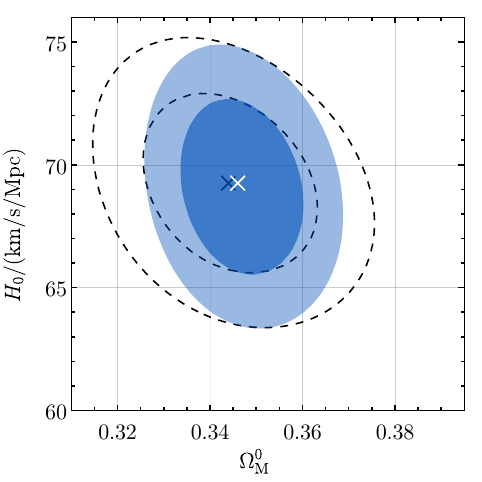} 
  \includegraphics[scale=1]{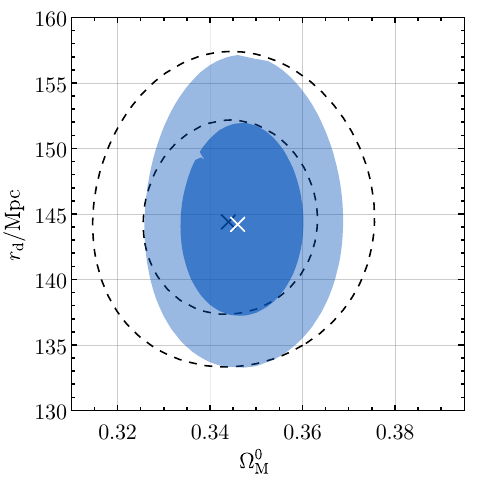}
  \includegraphics[scale=1]{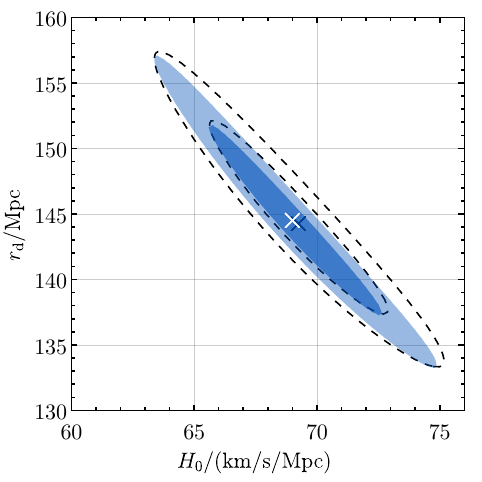} \hspace*{10em}
  \raisebox{10em}{\includegraphics[scale=1]{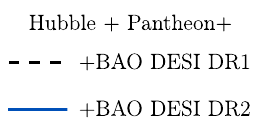}}
  \caption{Correlations of $H_0$--$\Omega_\text{M}^0$ (top left),
    $H_0$--$r_\text{d}$ (top right), $r_\text{d}$--$\Omega_\text{M}^0$ (bottom
    left) in the CKN model for the DESI BAO+Hubble+Pantheon+ DR1 (black dashed
    lines) and DESI BAO+Hubble+Pantheon+ DR2 (blue area) dataset at the $95\,\%$ and
    $68\,\%$ CL.}
  \label{fig:Corr_CKN_pantheon}
\end{figure}

\begin{figure}
    \includegraphics[scale=1]{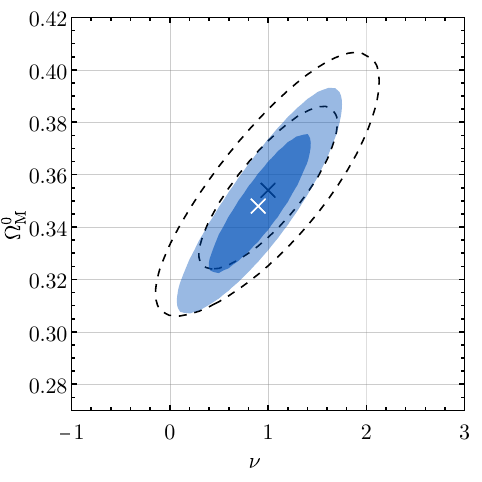} 
    \includegraphics[scale=1]{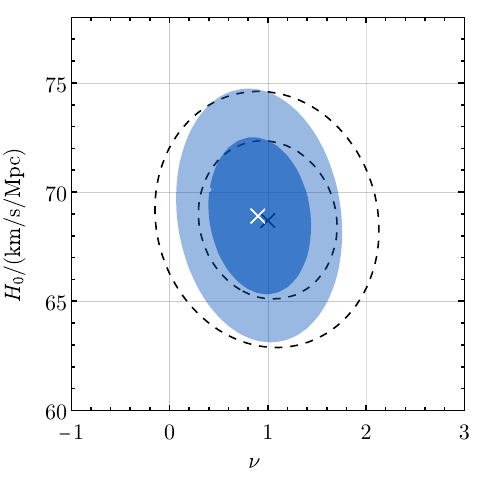}
    \includegraphics[scale=1]{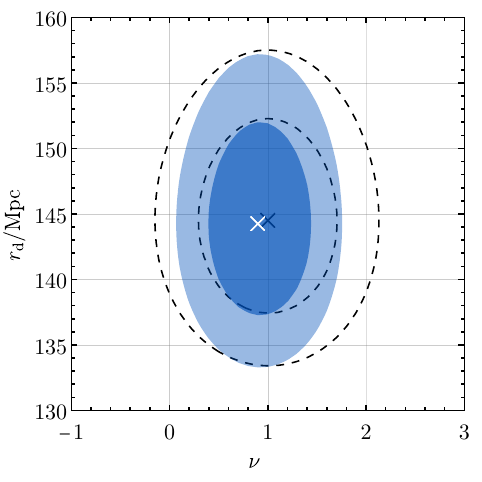} \hspace*{10em}
    \raisebox{10em}{\includegraphics[scale=1]{plots/legend_DESY5_DR2.pdf}}
    \caption{Correlations of $\nu$--$\Omega_\text{M}^0$ (top left),
      $\nu$--$H_0$ (top right), $\nu$--$r_\text{d}$ (bottom
      left) in the $\nu$CKN model for the DESI BAO+Hubble+DESY5 DR1 (black dashed lines)
      and DESI BAO+Hubble+DESY5 DR2 (blue area) dataset at the $95\,\%$ and $68\,\%$ CL.}
    \label{fig:Corr_nuCKN_DESY5}
\end{figure}

\begin{figure}
    \includegraphics[scale=1]{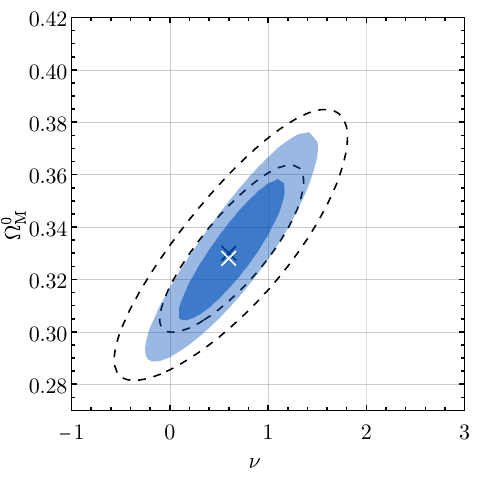} 
    \includegraphics[scale=1]{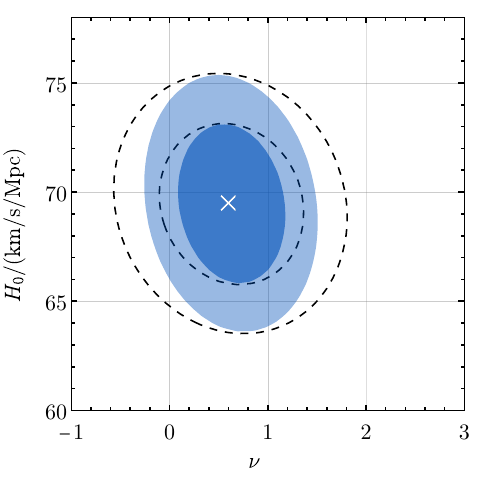}
    \includegraphics[scale=1]{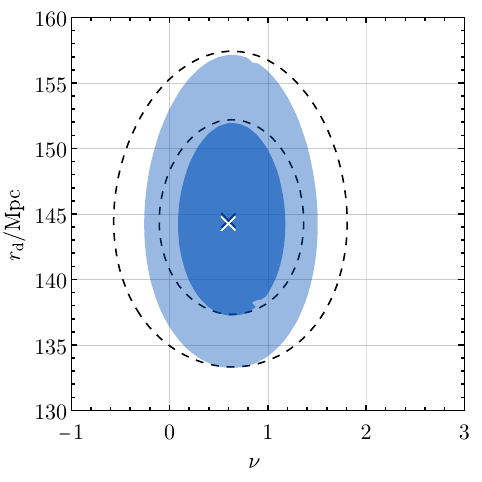} \hspace*{10em}
    \raisebox{10em}{\includegraphics[scale=1]{plots/legend_pantheon_DR2.pdf}}
    \caption{Correlations of $\nu$--$\Omega_\text{M}^0$ (top left),
      $\nu$--$H_0$ (top right), $\nu$--$r_\text{d}$ (bottom
      left) in the $\nu$CKN model for the DESI BAO+Hubble+Pantheon+ DR1 (black dashed
      lines) and DESI BAO+Hubble+Pantheon+ DR2 (blue area) dataset at the $95\,\%$ and
      $68\,\%$ CL.}
    \label{fig:Corr_nuCKN_pantheon}
\end{figure}

The improved statistics of the new measurements on the fit can also be seen in
the correlation plots of the CKN and $\nu$CKN models for both supernova datasets
in Figures~\ref{fig:Corr_CKN_DESY5}~to~\ref{fig:Corr_nuCKN_pantheon}, leading to
a smaller $95\,\%$ and $68\,\%$ confidence level (CL) area compared to the
results from DESI BAO DR1.

\begin{table}
    \caption{Results of the best-fit points for the CKN and $\nu$CKN case to the
        datasets of DESI BAO DR2 and Hubble, once combined with DESY5 and once with
        Pantheon+ data. Shown are the results for Hubble-today $H_0$, the matter
        density parameter $\Omega_\text{M}^0$, the drag epoch $r_\text{d}$, the
        parameter $\nu$ and the minimal $\chi^2_\text{min}$ over the degrees of
        freedom (DOF).}
    \begin{center}
    \begin{tabular}{l c c c c c} 
    \toprule
    {\bf Model}/Datasets & $H_0/(\text{km/s/Mpc})$ & $\Omega_\text{M}^0$ & $r_\text{d}/\text{Mpc}$ & $\nu$ & $\chi^2_\text{min}/\text{DOF}$\\
    \midrule
    \bf{CKN}\\
    + DESY5     & $68.83\pm2.35$ & $0.352\pm0.009$ & $144.27\pm4.85$ & -- & 1674/1871 \\
    + Pantheon+ & $69.09\pm2.36$ & $0.347\pm0.009$ & $144.23\pm4.85$ & -- & 1437/1632 \\
    \midrule
    \bf{$\bm{\nu}$CKN}\\
    + DESY5     & $68.90\pm2.38$ & $0.348\pm0.018$ & $144.26\pm4.85$ & $0.92\pm0.35$ & 1674/1870 \\
    + Pantheon+ & $69.46\pm2.40$ & $0.330\pm0.018$ & $144.21\pm4.85$ & $0.64\pm0.36$ & 1436/1631 \\
    \bottomrule
    \end{tabular}
    \label{tab:CKN_with_data}
  \end{center}
\end{table}

\begin{table}
  \caption{Results of the best-fit points from the $\Lambda$CDM, $\omega$CDM and
    $\omega_0 \omega_a$CDM model to the datasets of DESI BAO DR2 and Hubble, once
    with DESY5 and once with Pantheon+ data. Shown are the results for
    Hubble-today $H_0$, the matter density parameter $\Omega_\text{M}^0$, the drag
    epoch $r_\text{d}$, the parameters $\omega_0$ and $\omega_a$, and the minimal
    $\chi^2_\text{min}/\text{DOF}$.}
  \begin{center}
    \begin{tabular}{l c c c c c c } 
      \toprule
      {\bf Model} & $H_0$ in & $\Omega_\text{M}^0$ & $r_\text{d}$ in & $\omega$ or $\omega_0$ & $\omega_a$ & $\chi^2_\text{min}/\text{DOF}$ \\
      /Datasets   & $\text{km/s/Mpc}$ & & Mpc\\
      \midrule
      \textbf{$\bm{\Lambda}$CDM}\\
      + DESY5     & $69.77\pm2.38$ & $0.309\pm0.008$ & $144.28\pm4.85$ & -- & -- & 1681/1871  \\
      + Pantheon+ & $70.10\pm2.39$ & $0.303\pm0.008$ & $144.21\pm4.85$ & -- & -- & 1439/1632 \\
      \midrule
      \textbf{$\bm{\omega}$CDM}\\
      + DESY5     & $68.71\pm2.37$ & $0.297\pm0.009$ & $144.11\pm4.85$ & $-0.88\pm0.04$ & -- & 1670/1870 \\
      + Pantheon+ & $69.32\pm2.40$ & $0.297\pm0.008$ & $144.10\pm4.85$ & $-0.92\pm0.04$ & -- & 1435/1631 \\
      \midrule
      \textbf{$\bm{\omega_0 \omega_a}$CDM}\\
      + DESY5     & $68.69\pm2.50$ & $0.321\pm0.013$ & $143.70\pm5.01$ & $-0.78\pm0.07$ & $-0.77\pm0.39$ & 1668/1869 \\
      + Pantheon+ & $69.49\pm2.28$ & $0.302\pm0.017$ & $143.60\pm4.58$ & $-0.91\pm0.06$ & $-0.11\pm0.43$ & 1434/1630 \\
      \bottomrule
    \end{tabular}
    \label{tab:LCDM_with_data}
  \end{center}
\end{table}

\begin{table}
  \caption{Comparison between CKN and $\nu$CKN and alternative cosmological
    models. Shown is the difference $\Delta \chi^2 = \chi^{2,
    (\nu)\mathrm{CKN}}_\text{min} - \chi^{2, \mathrm{alt. model}}_\text{min}$ for
    both datasets and the difference $\Delta$AIC. The negative values indicate a
    preference of CKN and $\nu$CKN over the alternative models, respectively. See
    text for a comparison of the $\nu$CKN and $\Lambda$CDM model in terms of
    significances.}
  \begin{center}
    \begin{tabular}{c c c c c c} 
    \toprule
    Models & $\Delta \chi^2_\text{DESY5}$  & $\Delta \text{AIC}_\text{DESY5}$ & & $\Delta \chi^2_\text{Pantheon+}$  & $\Delta \text{AIC}_\text{Pantheon+}$\\
    \midrule
    \textbf{CKN with}\\
    $\Lambda$CDM           & $-6.90$ & $-6.90$ && $-2.05$ & $-2.05$ \\
    $\omega$CDM            & $3.14$  & $1.14$  && $2.26$  & $0.26$  \\
    $\omega_0 \omega_a$CDM & $5.74$  & $1.74$  && $2.43$  & $-1.57$ \\
    \midrule
    \textbf{$\bm{\nu}$CKN with}\\
    $\Lambda$CDM           & $-6.94$ & $-4.94$ && $-3.07$ & $-1.07$ \\
    $\omega$CDM            & $3.09$  & $3.09$  && $1.24$  & $1.24$  \\
    $\omega_0 \omega_a$CDM & $5.69$  & $3.69$  && $1.41$  & $-0.59$ \\
    \bottomrule
    \end{tabular}
    \label{tab:CKN_with_models}
  \end{center}
\end{table}

\begin{table}
    \caption{Difference of the $\chi_\text{min}^2$ values $\Delta
        \chi^2_{\text{DR2} - \text{DR1}}$ between the fit with DESI BAO DR1 and DR2
        combined with Hubble measurements and either the DESY5 or the Pantheon+
        supernova dataset for different cosmological models considered in this
        work. Additionally, we show the change in the difference between the
        $\chi_\text{min}^2$ values of the different models and the $\chi_\text{min}^2$ of the $\Lambda$CDM model $(\Delta
        \chi^{2})^{\Lambda\text{CDM}}$
        between the two data releases DR1 and DR2.}
    \begin{center}
    \begin{tabular}{l c c c c} 
    \toprule
    \multirow{2}{*}{Models} & \multicolumn{2}{c}{$\Delta \chi^2_{\text{DR2} -
    \text{DR1}}$} & \multicolumn{2}{c}{
      $ 
      (\Delta \chi^{2})^{\Lambda\text{CDM}}_\text{DR2} -
      (\Delta \chi^{2})^{\Lambda\text{CDM}}_\text{DR1}
      $
    }
    \\
    \cmidrule(l){2-3} 
    \cmidrule(l){4-5}
    & DESY5 & Pantheon+ & DESY5 & Pantheon+ \\
    \midrule
    CKN                    & $-2.85$ & $-2.84$ & $-2.34$ & $-0.90$ \\
    $\nu$CKN               & $-2.91$ & $-3.76$ & $-2.39$ & $-1.83$ \\
    $\Lambda$CDM           & $-0.52$ & $-1.93$ & --      & --      \\
    $\omega$CDM            & $-3.72$ & $-3.59$ & $-3.20$ & $-1.66$ \\
    $\omega_0 \omega_a$CDM & $-3.29$ & $-3.13$ & $-2.77$ & $-1.20$ \\
    \bottomrule
    \end{tabular}
    \label{tab:Change}
    \end{center}
\end{table}

% ######################################################################################
% #################################### Discussion ######################################
% ######################################################################################
\section{Discussion}
\label{sec:discussion}

As expected, the higher statistics coming with the DR2 leads to smaller
uncertainties on the model parameters and is visible in their corresponding
correlation plots for the $(\nu)$CKN models.
The widely discussed findings from DR1, that models featuring a time-varying
dark energy are preferred with respect to $\Lambda$CDM, is further strengthened
by the recent data release of the DESI experiment.
Remarkably, this trend is also visible in the ($\nu$)CKN models. For the
Pantheon+ dataset, the $\nu$CKN model experiences a larger improvement in the
fit between DR1 and DR2 compared to the CKN model.
The trend revealed through the new DESI data release fuels the hope that we
are moving towards a future where the mysteries of dark energy may at last begin
to unfold. 

% ######################################################################################
% #################################### Acknowledgements ################################
% ######################################################################################
\section*{Acknowledgements}
P.A.~is supported by the \textit{Stu\-di\-en\-stif\-tung des deutschen Volkes}.
M.H.~is supported by the Spanish
grants PID2023-147306NB-I00 and CEX2023-001292-S (MCIU/AEI/10.13039/501100011033),
as well as CIPROM/2021/054 (Generalitat Valenciana).

% ######################################################################################
% #################################### Bibliography ####################################
% ######################################################################################

\end{document}